\documentclass[article]{IEEEtran}

\title{Strongly Resilient Non-Interactive Key Predistribution
For Hierarchical Networks}

\author{Hao Chen}

\author{Hao Chen
 \thanks{Hao Chen is with the Software Engineering Institute, East
China Normal University, Shanghai 200062, China. EMAIL:
haochen@sei.ecnu.edu.cn }}

\begin{document}

\maketitle

\begin{abstract}

Key establishment is the basic necessary tool in the network
security, by which pairs in the network can establish shared keys
for protecting their pairwise communications. There have been some
key agreement or predistribution schemes with the property that the
key can be established without the interaction
(\cite{Blom84,BSHKY92,S97}). Recently the hierarchical cryptography
and the key management for hierarchical networks have been active
topics(see \cite{BBG05,GHKRRW08,GS02,HNZI02,HL02,Matt04}. ). Key
agreement schemes for hierarchical networks were presented in
\cite{Matt04,GHKRRW08} which is based on  the Blom key
predistribution scheme(Blom KPS, [1]) and pairing. In this paper we
introduce generalized Blom-Blundo et al key predistribution schemes.
These generalized Blom-Blundo et al key predistribution schemes have
the same security functionality as the Blom-Blundo et al KPS.
However different and random these KPSs can be used for various
parts of the networks for enhancing the resilience. We also present
key predistribution schemes from a family hyperelliptic curves.
These key predistribution schemes from different random curves can
be used for various parts of hierarchical networks. Then the
non-interactive, identity-based  and dynamic key predistributon
scheme based on this
generalized Blom-Blundo et al KPSs and hyperelliptic curve KPSs for hierarchical networks with the following properties are constructed.\\
1)$O(A_KU)$ storage at each node in the network where $U$ is the expansion number and $A_K$ is the number of nodes at the $K$-th level of
the hierarchical network;\\
2)Strongly resilience to the compromising of arbitrary many leaf and internal nodes;\\
3)Information theoretical security without random oracle.\\

\end{abstract}

\section{Introduction }

Key establishment is basic tool for secure communication in
networks, two nodes in networks can have agreed shared key that is
only known to them, thus allowing the shared key for protecting
their communications. In many environment there is significant
advantage to non-interactive key agreement schemes which need not to
use any communication between nodes. The Diffie-Hellman type key
agreement protocol(see \cite{BM03}) is non-interactive, but some
known public keys are needed  which is a impractical for large
networks. Recently  key agreement using key predistribution schemes
have
been presented for very large networks such as, hierarchical networks and wireless sensor networks(\cite{EG02,CPS03,Matt04,DDHVKK05,GHKRRW08,BST08}). \\

The key predistribution scheme(KPS) was proposed by R.Blom
 in Eurocrypt 84 (\cite{Blom84}). It was extended by C. Blundo et al in
Crypto 92 \cite{BSHKY92}. This cryptographic primitive has been a
basic ingredient in the security of wireless sensor networks(see
\cite{DF06,DDHVKK05}) and hierarchical  systems(see
\cite{Matt04,GHKRRW08}). However in the Blom and Blundo et al KPS,
the size of the finite field in the  KPS has to be larger than the
number of users. The unique form of Blom-Blundo et al KPS has no
flexibility in practical application. These are  real
drawbacks.  \\

In a HIERARCHICAL networks with $n$ nodes, the root authority only needs to distribute the secret information to a small number
of large organizations or group leaders, and then each of these can further distribute the secret information to smaller and smaller units(see \cite{Matt04,GHKRRW08}). In this way we can think the nodes are arranged on a tree, the root of tree distributes the secret information of its children nodes and then each of these distributes secret information to its children nodes... each node only get its secret information from its parent node. Finally the leaf nodes get their secret information from
their parent nodes. Each pair of nodes at the same level  (including the leaf nodes and internal nodes) can compute their shared key by the secret information and the identities of themselves and their parents. This would help for group level authentication and confidentiality in the whole hierarchical network. The expansion number $U$ is the maximal number of children nodes. \\

In the application  such as tactical networks, mobile ad-hoc networks, it is more reasonable to assume a {\em Hierarchical} network structure than a central trusted authority (see \cite{Matt04,RM05,GHKRRW08}). On the other hand, the using of Hierarchical network structure can reduce the workload of of the TAs. The Hierarchical identity based encryption (HIBE) was studied in \cite{HL02,GS02,BBG05}.  In \cite{CHK03}, HIBE was used for the construction of forward secure encryption. The hierarchical key agreement has been studied in \cite{Matt04,GHKRRW08}.\\

In previous constructed key agreement schemes in \cite{Matt04} and \cite{GHKRRW08}, every node in the hierarchical network needs the storage of $\frac{\prod{(t_i+1)(t_i+2)}}{2}$ elements of the base field for resisting the compromising  of $t_i$ nodes at $i$-th level of the hierarchical networks.  It will grows exponentially when the number of levels in the hierarchical network tends to the infinity. The KAS in \cite{GHKRRW08} can only resist the attack of compromising  arbitrary many leaf nodes. The security of KAS in \cite{GHKRRW08}was proved with the random oracle model.  The identity based key agreement scheme of \cite{GHKRRW08} is dynamic, nodes can be added at each level of the hierarchy without changing the information of other nodes.\\

In this paper we construct generalized Blom-Blundo et al key predistribution schemes and key predistribution schemes from a family of hyperelliptic curves. New random polynomials are introduced in the functions computing shared keys in these generalized Blom-Blundo et al key predistribution schemes. Hyerelliptic curve KPSs are constructed from different random curves. These new randomness and flexibility of our key predistribution schemes can be used to construct strongly resilient key predistribution schemes for hierarchical networks with low storage, communication and computation cost. The size of the base field of our new key predistribution schemes depends only on the expansion number $U$ of the hierarchical network and the storage of every node is $O(A_KU)$, where $A_K$ is the number of nodes at $K$-th level of the hierarchical network. Moreover the constructed hierarchical network key predistribution schemes are dynamic and non-interactive. Our key predistribution schemes for hierarchical networks can resist the compromising of arbitrary many of nodes with very low storage at every node.\\

\section{Blom-Blundo et al KPS}

Now we recall the definition of KPS by following the presentation in the
paper of Stinson \cite{S97}. Suppose we have a Trusted Authority (TA) and a set of users ${\bf
U}=\{1,...,n\}$. Let
$2^{{\bf U}}$  be the set of all subsets of the user set ${\bf U}$. ${\bf P} \subset
2^{{\bf U}}$ will denote the collection of all
privileged subsets to which the TA is distributing keys. ${\bf F}$
will denote the collection of all possible coalitions(forbidden subsets)
 against which
each key is remain secure. In the {\em Key Predistribution Scheme}, at
the set up stage, each user $i$ get its secret information $u_i$
from the TA, where $u_i$ is taken in a finite dimensional linear
space over $GF(q)$. Once the secret information $u_i$, $i=1,...,n$ ,
is given to each user, in the computation stage, for any privileged
subset $T \in {\bf P}$, the users in the privileged subset $T$ can
compute the shared key $k_T \in GF(q)$ for their communications. No
forbidden subset $J \in {\bf F}$ disjoint from $T$ can get any
information of the key $k_T$. This is called $({\bf P},{\bf
F})$-KPS. When  ${\bf P}$ consists of all subsets of ${\bf U}$
with $t$ elements and ${\bf F}$ consists of
subsets with at most $w$ elements, we call it $t$-variable and $w$-secure KPS. Thus a $t$-variable and $w$-secure
 KPS can be used to get the shared keys of
any subset with $t$ users, which is secure against the attack of any $w$ users.\\

Generally the KPS is required information theoretically secure against the attack of the coalition of users, for the  more formal
presentation we refer to  \cite{Blom84,BSHKY92,S97}.\\

The  secret information $u_i$, $i=1,...,n$, is in the finite
dimensional linear space over the finite field $GF(q)^h$, where $q$ is
a prime power. Thus the storage is $hlog_2(q)$ bits. The shared key $k_T$, for each privileged subset $T \in {\bf
P}$, is in $GF(q)$. In the computation stage, each user $i$ in $T$  computes
$k_T$ from its secret information $u_i$ and the IDs of other users in the set $T$. Only the arithmetic in
$GF(q)$ is involved. We call $GF(q)$ the base field of the KPS.\\

The first KPS proposed in \cite{Blom84} is a $2$-variable and $w$-secure KPS, and it
was generalized in \cite{BSHKY92} to a $t$-variable and $w$-secure KPS. Let $q$ be a prime power satisfying $ q \geq n$.
Each user $i$ is
assigned to an element $e_i \in GF(q)$ as its identity. The TA takes
a random $t$ variable symmetric polynomial in $GF(q)[x_1,...,x_t]$
of the form $f(x_1,...,x_t)=\Sigma_{j_1=1}^{w+1}\cdots
\Sigma_{j_t=1}^{w+1} a_{j_1 \cdots j_t} x_1^{j_1}\cdots x_t^{j_t} $ with coefficients
$a_{j_1 \cdots j_t}$ in
$GF(q)$ where $a_{j_1...j_t}=a_{j_{i_1}...j_{i_t}}$, that is, $f(x_1,...,x_t)=\Sigma_{j_1=1}^{w+1}\cdots
\Sigma_{j_t=1}^{w+1} a_{j_1 \cdots j_t} x_1^{j_1}\cdots x_t^{j_t} \in
GF(q)[x_1,...,x_t]$ and $f(x_1,...,x_t)=f(x_{i_1},...,x_{i_t})$(
$\{i_1,...,i_t\}$ is an arbitrary  permutation of
$\{1,...,t\}$). This polynomial is only known to the TA. The
symmetric $(t-1)$ polynomial  $f(e_i, x_2,...,x_t)$ is given to the
user $i$, $i=1,...,n$, as its secret information. For any privileged
subset $T=\{e_{i_1},...,e_{i_t}\}$, each user in this subset $T$ can
compute the shared key $k_T=f(e_{i_1},...,e_{i_t})$. \\

In the case $t=2$, this is just the KPS in \cite{Blom84}. The bit length
of secret information stored by each user in Blom-Blundo et al KPS is $\displaystyle{w+t-1 \choose t-1} \cdot log_2(q)$.\\

\section{Generalized Blom-Blundo et al key predistribution schemes}

In this section we present the generalized $2$-variable and $w$-secure Blom-Blundo et al KPS,
which can be extended easily to $t$-variable and $w$-secure  KPS.\\

Let $GF(q)$ be a fixed finite field, there are at least $\frac{q^t-\Sigma_{d|t}q^d}{t}$ distinct degree $t$ irreducible
polynomials in $GF(q)[x]$  Set $P(x)=p_1(x)  \cdots  p_h(x)$, where $p_i$'s are degree $t$ irreducible polynomial in $GF(q)[x]$.
This is a degree $H=ht$ polynomial in $GF(q)[x]$ which is not zero at any element in $GF(q)$.  Set $u(x)=\frac{f(x)}{P(x)}$,
where $f(x)$ is a degree $w$ polynomial. Because $P(x) \neq 0$ for any $x \in GF(q)$, thus $u(x)$ is defined for any $x \in GF(q)$.  Let $u_1=\frac{f_1}{P},...,u_{w+1}=\frac{f_{w+1}}{P}$, where $f_1,...,f_{w+1}$ is a base of the linear space of all polynomials in $GF(q)[x]$ with degree less than or equal to $w$,  be a base of the linear space of all these functions, for example $u_1(x)=\frac{1}{P(x)},u_2(x)=\frac{x}{P(x)},...,u_{w+1}(x)=\frac{x^{w}}{P(x)}$. \\

Suppose $H \geq w$ the $2$-variable and $w$-secure KPS associated with $P(x)$ on the set of $q$ users defined over $GF(q)$ can be constructed as follows.  The elements in $GF(q)$ are assigned to the users as their IDs. The TA takes a random $F(P,Q)=\Sigma_{i=1,j=1}^{w+1} a_{ij}u_i(P) u_j(Q)$, where $a_{ij}=a_{ji}$(then $F(P,Q)=F(Q,P)$) where $P,Q \in GF(q)$. The function $F(P=e_i,Q)$, as a function of $Q$, where $e_i \in GF(q)$, can be given to the user $e_i$ as its secret information. The shared key of the users with IDs $e_i$ and $e_j$ is $F(P=e_i,Q=e_j)$. The bit length of the secret information stored by each user is $(H+w+2)log_2(q)$. Here $(H+1)log_2 q$ bits are used to store the polynomial $P(x)$.\\

{\bf Theorem 1.} {\em Suppose $H \geq w$ the above KPS is $w$-secure.}\\

{\bf Proof.} We take the matrix of $w+1$ rows and $q$ columns with the entry at $i$ row and $j$ column is $u_i(x_j)$, where $x_j$ is the $j$-th element in $GF(q)$. This is actually a rank $w+1$ matrix. Actually any linear combination of  $w+1$ rows ${\bf v_{1}},...,{\bf v_{w+1}}$  of this matrix can not be zero at more than $w$ positions, since the function $c_1 u_1+ \cdots + c_{w+1}u_{w+1}=\frac{c_1f_1+\cdots+c_{w+1}f_{w+1}}{P}$ cannot have more than $w$ zero points. Then the $w$-security of the above KPS follows from the  same argument as in [1]. \\

The functions in the generalized Blom-Blundo et al KPSs have poles at the extension fields of $GF(q)$. If the polynomials $P$'s are distinct, these poles are distinct elements in the extension fields. Thus it is impossible for these  functions in $KPS(P_{random})$'s have an monic polynomial relation. That is, it is impossible to express the symmetric function used in one generalized Blom Blundo et al KPS as the polynomials of symmetric functions of other different generalized Blom Blundo et al KPSs. \\

The $t$-variable version of the generalized Bom-Blundo et al KPSs will  not be used in the hierarchical network key predistribution schemes given in section V. We include the construction here for the convenience of the readers. The $t$-variable and $w$-secure generalized Blom-Blundo et al KPS associated with $P(x)$ on the set of $q$ users defined over $GF(q)$ can be constructed as follows.  The elements in $GF(q)$ are assigned to the users as their IDs. The TA takes a random $F(P_1,...,P_t)=\Sigma_{i_1 \cdots i_t} a_{i_1\cdots i_t} u_{i_1}(P_1)\times \cdots \times u_{i_t}(P_t)$, where $a_{i_1 \cdots i_t}$ are symmetric about its subindices (then $F$ is symmetric about its variables) where $P_1,...,P_t \in GF(q)$. The $t-1$ variable function $F(P_=x,P_2,...,P_t)$ can be given to the user with $ID=x$ as its secret information. The shared key of the $t$ users with IDs $e_1,...,e_t$ is $F(e_1,...,e_t)$. The bit length of the secret information stored by each user is $\displaystyle{t+w-1 \choose t-1}log_2(q)+(H+1)log_2q$. Here $(H+1)log_2q$ bits are used for the storage of the polynomial $P(x)$.\\

The proof of the $w$-security of this $t$-variable KPS is directly since any $w+1$ columns of the matrix in Theorem 1 are linearly independent.\\

Then how many different such KPSs can we have?
We know there are at least ${\bf B_H}=\Sigma_{t|H}(\frac{q^t-\Sigma_{d|t}q^d}{t})^{\frac{H}{t}}$ polynomials $P(x)$
from the above argument corresponding to at least ${\bf B}$ such KPSs.
When $w$ is a prime number ${\bf B_H}=\frac{q^H-q}{H}$.
This is quite large when both $q$ and $H$ satisfying $q>H$ tends to the infinity.
Thus there are sufficiently such different $KPS(P)$'s for the randomness we need in the design of KPS
for the wireless sensor networks. Generally this number can be computed by zeta functions associated with the rational curve(see [16]).\\

When $f_1=1,...,f_{w+1}=x^{w}$ in the above generalized $2$-variable and $w$-secure KPS,
we have the shared key is computed by the function
$\Sigma_{i=0,j=0}^w, a_{ij} \frac{x^i}{P(x)} \frac{y^j}{P(y)}=\frac{\Sigma_{i,j=0}^w a_{ij}x^iy^j}{P(x)P(y)}$.
There are at least ${\bf B_H}=\Sigma_{t|H}(\frac{q^t-\Sigma_{d|t}q^d}{t})^{\frac{H}{t}}$
possible polynomials  $P(x) \in GF(q)[x]$ in the computation of the shared keys.
Hence the shared keys can be adjusted by these polynomials. So the randomness we needed in the design of KPS comes
from these polynomials $P \in GF(q)[x]$.\\

How can we use these irreducible polynomials in the implementation
of the generalized Blom-Blundo et al KPSs? From the theory of finite
fields, there are an enumeration of irreducible polynomials of
arbitrary fixed degree. For these low degrees,  some tables of
irreducible polynomials over $GF(2)$ and $GF(3)$ were listed in the
standard textbooks of finite fields. It can be used for the
implementation of
 generalized Blom-Blundo et al KPSs for which we take $h=\frac{w}{t}$  large positive integer and $t$ small positive integer.\\

{\bf Example 1.} Let $p(x)=1+2x+x^3 \in GF(9)[x]$. It is to check $p(x)$ is an irreducible polynomial in $GF(3)[x]$ and thus irreducible in $GF(9)[x]$, since the root is in $GF(27)$ and the intersection of $GF(9)$ and $GF(27)$ is $GF(3)$. Set $f_1(x)=\frac{1}{p(x)}, f_2(x)=\frac{x}{p(x)},f_3(x)=\frac{x^2}{p(x)},f_4(x)=\frac{x^3}{p(x)}$. We can have a $2$-variable and $3$-secure $KPS(p)$ on the set of $9$ players by taking random function $F(x,y)=\Sigma_{i=1,j=1}^4a_{ij} f_if_j=\Sigma_{i=0,j=0}^3  a_{ij} \frac{x^i}{p(x)} \frac{y^j}{p(y)}$, where $a_{ij}=a_{ji}$ are random elements in $GF(9)$. \\

{\bf Example 2.} Let $p(x)=x^7+x+1 \in GF(2)[x]$. This is an irreducible polynomial in $GF(2)[x]$. It is easy to check $p(x)$ is also irreducible in $GF(2^{11})[x]$, otherwise the intersection of $GF(128)$ and $GF(2^{11})$ is bigger than $GF(2)$. If $7h \leq 2^{11}=2048$, the functions $f_1=\frac{1}{p(x)^h}, f_2=\frac{x}{p(x)^h},...,f_{7h}=\frac{x^{7h}}{p(x)^h}$ can be used to get a $2$-variable and $7h$-secure generalized Blom-Blundo et al KPS. The setup server takes a random symmetric function $F(x,y)=\Sigma_{i=0}^{7h} a_{ij} \frac{x^i}{p(x)^h} \cdot \frac{y^j}{p(x)^h}$ where $a_{ij}=a_{ji}$ are random elements in $GF(2^{11})$. The setup server then predistributes $F(e_i, y)$ to the sensor node with $ID=e \in GF(2^{11})$ as its secret information. The shared key of two sensor nodes with IDs $e,e' \in GF(2^{10})$ is $F(e,e')$. This generalized Blom-Blundo KPS can be used for at most $2^{11}=1024$ sensor nodes. Since $7$ is a prime number $\frac{2^{7}-2}{7}=18$, we have at least $18$ distinct degree $7$ irreducible polynomials in $GF(2)[x]$. These polynomials are also irreducible in $GF(2^{11})[x]$. If $7h \leq 2048$, we can have at least $(18^{h}$ distinct $2$-variable and $7h$-secure KPSs on the set of $2048$ sensor nodes. All these distinct KPSs have the same security functionality as $2$-variable and $7h$-secure Blom-Blundo et al KPS. Thus these distinct generalized Blom-Blundo et al KPSs can be used for the various parts of the wireless sensor networks.\\

The generalized Blom-Blundo et al key predistribution schemes can be used for disigning strongly resilient  wireless sensor networks KPSs(see \cite{CWSN}).

\section{Random key predistribution schemes  from hyperelliptic curves}

\subsection{Key predistribution schemes from a family of hyperelliptic curves}

Let $q$ be an odd prime power, $X_a$ be the hyperelliptic curve $y^2=x^{q}+q+a$ defined over $GF(q^2)$, where
$a \in GF(q)$ is an arbitrary element in  $GF(q) \subset GF(q^2)$. The genus of this curve is $\frac{q-1}{2}$(see [14]).
For each $x \in GF(q^2)$, $x^q+x=Tr_{GF(q^2)/GF(q)}(x)$ is an element in $GF(q)$. Thus $x^q+x+a \in GF(q)$. It is easy to show that
each element in $GF(q) \subset GF(q^2)$ is a square element, thus we have $2q^2$ affine $GF(q^2)$ rational points on $X_a$,
and one $GF(q^2)$ rational point $Q$ at the infinity. $x$ has a $2$-th pole at the point $Q$ and
$y$ has a $q$-th pole at the infinity. Let $L(uQ)$ be the linear space of rational functions on the hyperelliptic
curve with only pole at the point $Q$ and the pole order not bigger than $u$. It is known that $\{x^iy^j|2i+qj \leq u\}$,
under the reduction $y^2=x^q+x+a$, is a base of the function space $L(uQ)$ if $u \geq 2g-1=q-2$, which is a $u-g+1$ dimensional space over $GF(q)$. For example when $u=2q$, then $\{1,x,...,x^{\frac{q+1}{2}},y,yx,...,yx^{\frac{q-1}{2}}\}$ is a base of $L((2q)Q)$(see [14]). \\

Suppose $q \geq 5$. We have a key predistribution scheme over $GF(q^2)$ on the set of $2q^2$ users, the TA can take $X_a$ for a random $a \in GF(q)$ and
a random function $F(P_1,P_2)=\Sigma_{i,j=1}^{w+\frac{q+1}{2}} a_{ij} f_i(P_1) f_j(P_2) \in L((w+q-1)Q) \otimes L((w+q-1)Q)$, where $(P_1, P_2) \in X_a \times X_a$. Here $a_{ij}$ is symmetric about $i$ and $j$, $f_1,...,f_{w+\frac{q+1}{2}}$ is a base of $L((w+q-1)Q)$ of the form $x^h_1 y^{h_2}$. Then $F(P_1=W,P_2) \in L((w+q-1)Q)$ is given to the user with the $ID=W$ as its secret information. For the users with $ID=W$ and $ID=W'$, the shared key between them is $F(W,W') \in GF(q^2)$. It is clear that in this $(2,w)$ KPS over $GF(q^2)$ on the set of $2q^2$ users the storage of secret information of each user is $2(w+\frac{q-1}{2}) log_2(q)$ bits.\\

{\bf Theorem 2.} {\em The above key predistribution scheme is $w$-secure.}\\

{\bf Proof.} We consider the $(w+\frac{q+1}{2}) \times (2q^2)$ matrix by evaluating the $w+\frac{q+1}{2}$ base functions of $L((w+q-1)Q)$ at the $2q^2$ points described as above. This is actually the generator matrix of the algebraic geometric code(see [14]). It is well-known the minimum Hamming distance of the dual code is at least $w+2$(see [14]). Thus any $w+1$ columns of the above matrix are linear independent vectors in $GF(q^2)^{w+\frac{q+1}{2}}$. From the construction of Blom key predistribution scheme in [1](also see \cite{DDHVKK05} pages 236-237), the above construction is a $w$-secure key predistribution scheme on $2q^2$ users.\\

In this family of key predistribution schemes $KPS(a)$ on the set of $2q^2$ users, where $a$ is the parameter of curve equation,  the shared keys are computed in a field with $q^2$ elements. The randomness of of these KPSs are from random curves instead of polynomials in the generalized Blom  KPSs.\\

Though we need not to use the $t$-variable case in section V for the key predistribution schemes of hierarchical networks the construction is included here for the convenience of the readers. The above $2$-variable and $w$-secure KPS can be extended to $t$-variable and $w$-secure KPS as follows. the TA can take $X_a$ for a random $a \in GF(q)$ and
a random function $F(P_1,,...,P_t)=\Sigma_{i_1...i_t=1}^{w+\frac{q+1}{2}} a_{i_1...i_t} f_{i_1}(P_1)\times \cdots \times f_{i_t}(P_{i_t}) \in L((w+q-1)Q) \otimes \cdots \otimes L((w+q-1)Q)$, where $(P_{i_1},..., P_{i_t}) \in X_a \times \cdots \times  X_a$. Here $a_{i_1...i_t}=a_{j_1..j_t}$, where $j_1...j_t$ is an arbitrary permutation of $i_1...i_t$, and $f_1,...,f_{w+\frac{q+1}{2}}$ is a base of $L((w+q-1)Q)$ of the form $x^h_1 y^{h_2}$. Then $F(P_{i_1}=W,P_{i_2},...,P_{i_t})$ of $t-1$ variables is given to the user with the $ID=W$ as its secret information. For the users with $ID_1=W_1,...,ID_t=W_t$, the shared key for them is $F(W_1,...W_t) \in GF(q^2)$. It can be proved similarly as above  that  this $t$-variable and $w$-secure KPS over $GF(q^2)$ on the set of $2q^2$ users.  The storage of secret information of each user is $2\displaystyle{t+w+\frac{q-3}{2} \choose t-1}log_2(q)$ bits. The detailed construction and the proof will be included in our future paper [9].\\

\subsection{Implementation}

In the key predistribution schemes from hyperelliptic curve $X_a$ where $a$ can take any element in $GF(q)$, the TA  can assign the coordinates of the $GF(q)$ rational points of the hyperelliptic curve $X_a, a \in GF(q)$  to the $2q^2$ users as their IDs. Then the TA can fix a base of  the function space $L((w+q-1)Q)$ as above. The process of these key predistribution schemes is the same as in Blom KPS, the only difference is the polynomials and the elements of the finite field are replaced by rational functions in $L((w+q-1)Q)$ and $GF(q^2)$ rational points of the curve. It should be noted that the same monimial base as above can be used for arbitrary curve $X_a, a \in GF(q)$, in the process of the computation of the shared keys, the reduction used on the curve $X_a$ is $y^2=x^q+x+a$. The parameter $a$ playes the critical role in the computation of shared keys in the hyperelliptic curve key preditribution schemes.  Here  $(w+\frac{q+1}{2})log_2(q)$ bits of secret information need to be stored by each user.\\

\section{Strongly resilient key predistribution schemes for hierarchical networks}

Let $R$ be the root authority, it has at most $A_1$ children nodes $R_1$,...,$R_{A_1}$, each $R_i$ has $A_{(i)}$ children nodes, $R_{i1}...R_{iA_{(i)}}$. $A_2=\Sigma A_{(i)}$ is number of all nodes at the 3rd level. We assume the hierarchical system has $L+1$ levels. The node at the $K$ level is denoted by
$R_{i_1i_2..i_{K-1}}$, which has $A_{(i_1i_2...i_{K-1})}$ children nodes. Here $i_j$ is its number at the $j$-th level. Let $A_K=\Sigma A_{(i_1i_2...\i_{K-1})}$ is the number of all nodes at the $K+1$-th level. We assume $A_{(i_1...i_K)} \leq U$ for any possible subindices, that is, for each node, it has at most $U$ children nodes. $U$ is called the expansion number. \\

\subsection{Generalized Blom-Blundo et al key predistribution schemes for hierarchical networks}

We fix a prime power $q \geq 2U$ and a positive integer $t$ such that $\frac{q^t-\Sigma_{d|t}q^d}{t} \geq q$ and $2U-1=th$ for some positive integer $h$. We consider the $q$ irreducible polynomials of degree $t$ $P_1,...,P_{q} \in GF(q)[x]$ and a one-to-one correspondence  between $P_{\alpha_1},...,P_{\alpha_q}$ and the elements $\alpha_1,...,\alpha_q$ of $GF(q)$ will be used. For any parent node $R_{i_1...i_{K-2}}$ at the $K-1$-th level, each child node $R_{i_1...i_{K-2}j}$ at the $K$ level is assigned an element in $GF(q)$ as its ID. There are at least $(\frac{q^{t}-\Sigma_{d|t} q^d }{ t} )^h>q$ different $(2,2U-2)$ curve-KPS on the set of $2U$ users defined over $GF(q)$. The KPS associated with the polynomial $P_{\alpha_i}^h$ is denoted by $KPS(P_{\alpha_i})$\\

The root authority $R$ uses the random $KPS(P_s)$, where $s$ is a random element in $GF(q)$, to give the secret information to each of its child node $R_i$, where $ i \leq A_1$. The bit length of the secret information is $2(U-1) log_2(q)$. For each node $R_i$ at the 2nd level,  $R_i$ randomly picks up $KPS(P_{s_i})$, where $s_i \in GF(q)$ is random element in $GF(q)$, to give each of its child node the secret information. For any two $R_{i_1j_1}$ and $R_{i_2j_2}$ at the 3rd level, $R_{i_1}$ and $R_{i_2}$ at the 2nd level can have a shared key  $s_{i_1i_2}$ in $GF(q)$ from the $KPS(P_s)$, then $R_{i_1}$ and $R_{i_2}$ use $KPS(P_{s_{i_1i_2}})$ to give secret information to their children nodes $R_{i_1j}$'s and $R_{i_2j}$'s.  When $R_{ij_1}$ and $R_{ij_2}$ want to find their shared key, they can use $KPS(P_{s_i}$, and when $R_{i_1j_1}$ and $R_{i_2j_2}$ want to find their shared key, they can use $KPS(P_{s_{i_1i_2}})$. This process can proceed to all the levels. That is, $R_{i_1...i_w}$ randomly picks up $KPS(P_{s_{i_1...i_w}})$ for the shared key among its children nodes, and $R_{i_1...i_w}$ and $R_{i_1'...i_w'}$ use their shared key $s_{i_1i_1'...i_wi_w'}$ to fix a $KPS(P_{s_{i_1i_1'...i_wi_w'}})$, then this KPS is used for the shared key between the children nodes of $R_{i_1...i_w}$ and $R_{i_1'...i_w'}$.\\

The bit length stored in each node at the $K+1$-th level is $2A_K(2U-1)log_2(q)$ and the computation of the shared key is mainly the  $(2U-1)$ times of multiplications of the finite field $GF(q)$.\\

\subsection{Hyperelliptic curve key predistribution schemes for hierarchical networks}

We denote the generalized $(2,2U-2)$ curve-KPS on the set of $2U \leq 2q^2$ users defined over $GF(q^2)$ from the hyperellptic curve $X: y^2=x^q+x+a$ as in section 3.1 as $KPS(a)$ with parameter $a$ from the finite field $GF(q)$. We take a finite field
$GF(q^2)$ satisfying $2U \leq 2q^2$. The root authority $R$ uses the random $KPS(a)$, that is $a$ is randomly picked up from the finite field $GF(q)$, to give the secret information to each of its child node $R_i$. The bit length of the secret information is $2(2U+\frac{q-3}{2}) log_2(q^2)$. For each node $R_i$ at the 2nd level, $R_i$ randomly picks up $KPS(P_{a_i})$, where $a_i \in GF(q)$,  to give each of its child node the secret information. For any two $R_{i_1j_1}$ and $R_{i_2j_2}$ at the 3rd level, $R_{i_1}$ and $R_{i_2}$ at the 2nd level can have a shared key  $s_{i_1i_2}$ in $GF(q^2)$ from the $KPS(a)$, then $R_{i_1}$ and $R_{i_2}$ use $KPS(P_{s_{i_1i_2}^{q+1}})$, to give secret information to their children nodes $R_{i_1j}$'s and $R_{i_2j}$'s. It should be noted $s_{i_1i_2}^{q+1} \in GF(q)$ since $s_{i_1i_2} \in GF(q^2)$.  When $R_{ij_1}$ and $R_{ij_2}$ want to find their shared key, they can use $KPS(P_{a_i}$, and when $R_{i_1j_1}$ and $R_{i_2j_2}$ want to find their shared key, they can use $KPS(P_{s_{i_1i_2}^{q+1}})$. This process can proceed to all the levels. That is, $R_{i_1...i_w}$ randomly picks up $KPS(P_{s_{i_1...i_w}})$ , where $s_{i_1...i_w} \in GF(q)$, for the shared key among its children nodes. The nodes $R_{i_1...i_w}$ and $R_{i_1'...i_w'}$ use their shared key $s_{i_1i_1'...i_wi_w'}$ to fix a $KPS(P_{s_{i_1i_1'...i_wi_w'}^{q+1}})$, then this KPS is used for the shared key between the children nodes of $R_{i_1...i_w}$ and $R_{i_1'...i_w'}$.\\

The field size in this hyperelliptic  curve-KAS for the hierarchical system has to satisfy $q^2 \geq \frac{U}{2}$, which is much weaker than the previous KAS.\\

The bit length of the secret information stored in each node at the $K+1$-th level is $2A_K(2U+\frac{q-3}{2}) log_2(q^2)$ and at most $4U+q-3$ times of multiplications of the field $GF(q^2)$ are used for computing the shared key.\\

\subsection{Key predistribution  schemes for dynamic hierarchical networks}

In the above hierarchical KAS, when $q \geq 2A_{(i_1...i_{K-1})}$ is valid in genus 0 KPS and $q^2 \geq A_{(i_1...i_{K-1})}$ in hyperelliptic curve KPS, nodes can be added by the parent node $R_{i_1...i_{K-1}}$ to the hierarchy. That is, if we choose $q$ with suitable large size, the hierarchical nodes can added by the parent node without change the settings of other nodes.

\section{Information theoretical security}
Because the number of children nodes of each node $A_{(i_1i_2...i_K)} \leq U$ and we use $(2,2U-2)$ KPS, the adversary  compromising less than $2U$ nodes cannot get the full information of the KPS used, if the adversary compromise all children nodes (at the $K+1$-th level) of the nodes $R_{i_1...i_{K-1}}$ and $R_{i_1'...i_{K-1}'}$, the KPS used can be deleted and all the children nodes in the further levels of the nodes $R_{i_1...i_{K-1}}$ and $R_{i_1'...i_{K-1}'}$ and themselves can be deleted without any impact on the key agreement scheme of the other nodes, since we use the {\bf RANDOM} KPS associated with random polynomials or from random curves for the key predistribution for the un-compromised nodes and their children nodes. The point here is, after deleting the compromising nodes, their children nodes and their parent nodes, the secret information stored in un-compromised nodes is random and the shared keys of the un-compromised nodes are uniformly distrubited random variables from the view of  the compromised nodes.\\

\section{Conclusion}
In this paper the generalized Blom-Blundo eta la key predistribution schemes and key predistribution schemes from hyperelliptic curves have been constructed. This kind
of KPSs is flexible and can be used to construct hierarchical network key predistribution schemes. The size of shared keys only depends on the expansion numbers of nodes. These hierarchical network KPSs are identity based and dynamic. They are more efficient than the previously known hierarchical key agreement schemes and information theoretical secure against the compromising of arbitrary many internal and leaf nodes. The storage of each node is linear about the number of nodes at each level.\\

{\bf Acknowledgment:} The work was supported by the National
Natural Science Foundation of China Grant 10871068.\\

\end {document}